\documentclass{article}
\usepackage{csquotes}
\usepackage[utf8]{inputenc}
\usepackage[backend=biber,style=alphabetic]{biblatex}
\usepackage[english]{babel}
\usepackage[margin=1in]{geometry}
\usepackage{amsmath,amssymb,amsfonts,amsthm,mathtools}
\usepackage{braket} 
\usepackage{xcolor}
\usepackage{tikz}
\usetikzlibrary{fit}
\usepackage{quantikz}
\usepackage[inline]{enumitem}

\usepackage{hyperref}
\usepackage[nameinlink]{cleveref}

\hypersetup{colorlinks=true,linkcolor=blue,filecolor=magenta,urlcolor=cyan}
\newtheorem{theorem}{Theorem}[section]
\newtheorem{corollary}[theorem]{Corollary}
\newtheorem{lemma}[theorem]{Lemma}
\theoremstyle{remark}

\theoremstyle{definition}

\newtheorem{prop}{Proposition}

\newcommand{\R}{\mathbb{R}}
\newcommand{\C}{\mathbb{C}}

\title{Slices for normal forms in quantum informatiion and a Riccati-type equation}
\author{Ranee Brylinski\textsuperscript{*} and Jean-Luc Brylinski}
\date{November 2026}

\begin{document}

\maketitle

\section{Introduction}  We use the fact a slice to an orbit of a compact Lie group acting linearly meets each group orbit to derive known normal forms for pure and mixed qudits. We find the normal form from the gradient flow of a distance function, and observe this flow is governed by a matrix valued  ODE of Riccati type.

\section{The slice theorem}

Palais \cite{pa} proved in 1961 that for an orbit $\mathcal O$ of a compact Lie group $G$ acting smoothly on a smooth manifold $M$ and a point $x$ in $\mathcal O$ , there exists a locally closed submanifold $S \subset  M$ containing $x$, called \underline{slice} at x, such that:

(1) $S$ is invariant under the action of the isotropy  group $G_x$ of $x$.

(2) $S$ is transverse to the orbit $\mathcal O$ at $x$, i.e. the tangent space to $M$ at $x$ is the direct sum  of the tangent spaces spaces to $\mathcal O$ and to $S$.

(3) there is neighborhood $V$ of $x$ in $S$ such he transform $g\cdot V$ of $V$ under $g\in G$ does not intersect $V$ if $g$ does not belong to $G_x$.
\vskip .13 in

Then Palais proves that for such a slice $S$ and such a neighborhood $V$ of $x$ in $S$, the contracted product $G\times_{G_x}V$ is $G$-equivariantly   diffeomorphic to an open neightborhood $U$
 of $\mathcal O$ in $M$. More precisely, the contracted product $G\times_{G_x}V$ is the quotient of the product $G\times V$ by the action of $G_x$ where $g\cdot (h,y)=(hg^{-1}, g\cdot y)$. This is the total space of a $G$-equivariant fiber bundle over $\mathcal O=G/G_x$ with fiber $V$. The diffeomorphism $G\times_{G_x} V\to U$ maps $(g,y)$ to $g\cdot y$.
\vskip .14 in
We are interested in the case where $M=\R^n$ and the action of $G$ on $\R^n$ is linear. In that case we can assume, after a linear change of basis in $\R^n$, that $G$ acts by orthogonal matrices, and Palais proved that a suitable slice is the normal space $N_x\mathcal O$ to $\mathcal O$ in $M$. We denote by $\langle x\mid y\rangle$ the Euclidean dot product of vectors $x$ and $y$ in $\R^n$. We let $\mathfrak g$ be the Lie algebra of $G$, which acts infinitesimally on $\R^n$ by the skew linear transformation
\[
  \xi\cdot y = \left.\frac{d}{dt}\,[\exp(t\xi)\,y]\right\rvert_{t=0}.
\]
The tangent space to the orbit $\mathcal O$ at $x$ is $\mathfrak g\cdot x$; its orthogonal $(\mathfrak g\cdot x)^\perp$ is the normal space $N_x\mathcal O$.
\vskip .12in
We then have the following well-known theorem.

\begin{theorem}\label{thm:normal}
The normal space $N_x\mathcal O$ meets every $G$-orbit in $\R^n$.
\end{theorem}

\begin{proof}
Let $\mathcal O' = G\cdot y$ be another orbit. Intuition strongly suggests that a point in $\mathcal O'$ closest to $x$ should belong to the slice $S$. We formalize this by introducing the smooth distance function
\[
  F(g) = \lVert g\cdot y - x\rVert^2.
\]
Say this function achieves its minimum at $g$. Setting $z=g\cdot y$, we then have for any $\xi\in\mathfrak g$:
\begin{equation}
\left.\frac{d}{dt}\,\lVert \exp(t\xi)\cdot z - x\rVert^2\right\rvert_{t=0}=0.
\end{equation}
This derivative is equal to $-2\langle \xi\cdot z\mid x\rangle=2\langle z\mid \xi\cdot x\rangle$. Hence $z$ is orthogonal to $\mathfrak g\cdot x$, so $z\in N_x\mathcal O$.
\end{proof}
 
\section{Normal form for mixed $2$-qubits}

Consider a mixed state $\rho$ of $2$ qubits, where the first qubit belongs to Alice and the second belongs to Bob. We can form the reduced density matrices $\rho_A$ and $\rho_B$. The difference $i\bigl(\rho-\rho_A\otimes\rho_B\bigr)$ belongs to the tensor product $\mathfrak{su}(2)_A\otimes\mathfrak{su}(2)_B\subset \mathfrak{u}(2)_A\otimes\mathfrak{u}(2)_B=\mathfrak{u}(4)$, where $\mathfrak{u}(2)$ is the $4$-dimensional real vector space of skew-Hermitian operators on $\C^2$, and $\mathfrak{su}(2)$ is the $3$-dimensional subspace of traceless skew-Hermitian operators. The rescaled Pauli matrices $iX,iY,iZ$ form a basis of $\mathfrak{su}(2)$, and $iI,iX,iY,iZ$ form a basis of $\mathfrak{u}(2)$.

We want to find a normal form for the mixed state $\rho$ up to the symmetry group of local (i.e. one-qubit) gauge operators $G=SU(2)_A\times SU(2)_B$. We are thus led to consider the conjugation action of $G$ on the real vector space $i\cdot\bigl(\mathfrak{su}(2)_A\otimes\mathfrak{su}(2)_B\bigr)$. The infinitesimal action of the Lie algebra $\mathfrak{su}(2)_A\oplus\mathfrak{su}(2)_B$ on $V$ is given by Lie brackets of operators, i.e.
\begin{equation}
(U,V)\cdot (R,S)=([U,R],[V,S])
    \end{equation}

We will pick a $G$-orbit $\mathcal O$ in $V$ of maximal dimension, namely $dim(G)=6$. We set
\begin{equation}
\tau= i\bigl(a\, X\otimes X + b\, Y\otimes Y + c\, Z\otimes Z\bigr)
\end{equation}
for real numbers $a,b,c$ to be determined later. Let $\mathcal O$ be the orbit of$\tau$.

\begin{lemma}
The tangent space to $\mathcal O$ at $\tau$ is six-dimensional and has the basis
\begin{equation}
i X\otimes Y,i  X\otimes Z, i Y\otimes X, i  Y\otimes Z, i Z\otimes X, i Z\otimes Y
\end{equation}
assuming that $a^2,b^2$ and $c^2$ are non-zero and are all disjoint. 
\end{lemma}
\begin{proof}
The action of $iZ$ by brackets on $\mathfrak{su}(2)$ maps $iX$ to $2iY$ and $iY$ to $-2iX$. Hence the action of $(iZ,0)\in \mathfrak{su}(2)_A\oplus\mathfrak{su}(2)_BS$ maps $iX\otimes X$ to $iY\otimes X$ and $iY\otimes Y$ to $-iX\otimes Y$, hence it maps $\tau$ to $i\bigl(aY\otimes X-bX\otimes Y\bigr)$. Similarly the action of $(0,iZ)$ maps $\tau$ to $i\bigl(aX\otimes Y-bY\otimes X\bigr)$. Since $a\neq 0$, $b\neq 0$ and $a^2\neq b^2$, $iX\otimes Y$ and $iY\otimes X$ are real-linear combinations of these two tangent vectors, so both $iX\otimes Y$ and $iY\otimes X$ belong to the tangent space to $\mathcal O$ at $\tau$.

Similarly, using the elements $i[X,0], i[0,X],i[Y,0],i[0,Y]$ of $su(2)_A\oplus su(2)_B$ we find that $iY\otimes Z,iZ\otimes X,i iX\otimes Z,iZ\otimes X$ all belong to the tangent space. We have found $6$ linearly independent elements of the tangent space $T_{\tau}\mathcal O$; since the dimension of the tangent space if $\leq dim(G)=6$, these six elements must form a basis of $T_{\tau}\mathcal O$.
\end{proof}

From this we see that the normal space to $\mathcal O$ at $\tau$ is spanned by $iX\otimes X$, $iY\otimes Y$, and $iZ\otimes Z$. Hence every $G$-orbit in $V$ contains an element of the form $i\bigl(\alpha X\otimes X+\beta Y\otimes Y+ \gamma Z\otimes Z\bigr)$. Applying this to $i\bigl(\rho-\rho_A\otimes \rho_B\bigr)$ for a mixed $2$-qubit state $\rho$, we get the normal form theorem of Bloch and Fano\cite{fa}.

\begin{theorem}[Bloch--Fano normal form]
Any mixed $2$-qubit state is equivalent under the action of the local gauge  group $SU(2)_A\times SU(2)_B$ to a state of the form
\begin{equation}
\sigma=\sigma_A\otimes \sigma_B+\alpha X\otimes X+\beta Y\otimes Y+ \gamma Z\otimes Z.
\end{equation}
\end{theorem}

\section{The Schmidt decomposition}

Here we are considering the complex vector space $(\C^d)^{\otimes 2}$ of pure states of $2$ qudits. We denote by $\langle\phi |\psi\rangle $ the Dirac bracket, which is linear in the ket $|\psi\rangle$. We denote by $\langle \phi| \psi \rangle_{\R}$ its real part, which is a real-valued positive-definite inner product. In terms of the expansion $\phi=\sum_{\bar j \in [1,d]^2} \phi_{\bar j}|j_1j_2\rangle$ in the computational basis $(\bar j)$, we have

\begin{align*}
\langle \phi\mid\psi\rangle &= \sum_{\bar j}\overline{\phi_{\bar j}}\,\psi_{\bar j} \\
\langle \phi\mid\psi\rangle_{\R} &= \sum_{\bar j}(\Re(\phi_{\bar j})\,\Re(\psi_{\bar j}) + \Im(\phi_{\bar j})\,\Im(\psi_{\bar j}))
\end{align*}

The local group $SU(d)^2$ acts by complex-linear transformations on $(\C^d)^{\otimes 2}$ by
\[
(g_1,g_2)\cdot (|\phi_1\rangle\otimes|\phi_2\rangle)= (g_1|\phi_1\rangle)\otimes (g_2|\phi_2\rangle).
\]
We will consider the $G$-orbit $\mathcal O$ of the state $|\phi\rangle =c_1|11\rangle+\cdots+c_d|dd\rangle$.

\begin{lemma}
The tangent space to the orbit $\mathcal O$ at $|\phi\rangle$ is equal to
\begin{equation}
\left(\bigoplus_{j=1}^d \R\cdot i|jj\rangle\right)\;\oplus\;\left(\bigoplus_{j\neq k}\C\cdot |jk\rangle\right),
\end{equation}
assuming that each $c_j$ is a nonzero real number and that the $c_j^2$'s are all distinct.
\end{lemma}

\begin{proof}
The tangent space at $|\phi\rangle$ is $\mathfrak g\cdot|\phi\rangle$, where $\mathfrak g=\mathfrak{su}(d)_A\oplus\mathfrak{su}(d)_B$.
The Lie algebra $\mathfrak{su}(d)$ is generated by the following types of elements:
\begin{enumerate}
\item the diagonal elements $i |j\rangle\langle j|$ for $j=1,\dots,d$,
\item the elements $i\bigl(|j\rangle\langle k|+|k\rangle\langle j|\bigr)$ for $j\neq k$,
\item the elements $|j\rangle\langle k|-|k\rangle\langle j|$ for $j\neq k$.
\end{enumerate}

The first type of elements on Alice's side map $|\phi\rangle$ to $ic_j\,|jj\rangle$, so $i |jj\rangle$ belongs to the tangent space (and similarly on Bob's side).

The second type of elements on Alice's side map $|\phi\rangle$ to $i\bigl(c_k|jk\rangle+c_j|kj\rangle\bigr)$, and the second type on Bob's side map $|\phi\rangle$ to $i\bigl(c_j|jk\rangle+c_k|kj\rangle\bigr)$. Their span is $i \R\cdot|jk\rangle\oplus i \R \cdot|kj\rangle$ as long as $c_j^2\neq c_k^2$.

The third type of elements on Alice's side map $|\phi\rangle$ to $c_k|jk\rangle-c_j|kj\rangle$, and on Bob's side to $c_j|jk\rangle-c_k|kj\rangle$, which  span $\R\cdot|jk\rangle\oplus\R\cdot|kj\rangle$.
Putting these together gives the lemma.
\end{proof}

This implies the normal space $(\mathfrak g\cdot|\phi\rangle)^\perp$ to the orbit is the real span of the $|jj\rangle$. Hence we get the Schmidt normal form (Schmidt decomposition) (see \cite{ni-ch}).

\begin{prop}
Every pure $2$-qudit state is locally (i.e., under the $G$-action) equivalent to a state of the form $c_1|11\rangle+\cdots+c_d|dd\rangle$ where the $c_j$'s are  real numbers.
\end{prop}

 \begin{corollary}
 Every $2$-qudit state $|\phi\rangle$ admits the Schmidt decomposition
\begin{equation}
|\phi\rangle=\sum_{j=1}^dc_j |e_j\rangle\otimes |f_j\rangle 
\end{equation}
where the $|e_j\rangle$ and the $|f_j\rangle$ are ON bases of $\C^d$, and the $c_j$'s are non-negative real numbers.
\end{corollary}

\section{Implementing finite symmetries into the normal form}

We will work in the general situation of a compact Lie group $G$ acting by orthogonal linear transformations on a real vector space $V$ equipped with a positive-definite inner product $\langle .|.\rangle$. We wish to consider the case where we have a finite symmetry group $H$ acting by orthogonal linear  transformations on $V$ For example, when $V=su(2)\otimes su(2)$ and $G=SU(2)\times SU(2)$, $H$ would be the order  $2$ group generated by the flip of the two qubits.

The subspace $V^H$ fixed by $H$ is stable under the action of the subgroup $G^H$ of $G$ of elements which commute with $H$. Then we know from Theorem \ref{thm:normal} that the $G^H$-orbit of a vector $z$ in $V^H$ meets the normal space $(\mathfrak g^H\cdot z)^\perp$ to the $G^H$-orbit of $z$ inside $V^H$. The orthogonal space here is taken inside $V^H$. We have

\begin{lemma}
The orthogonal space to $\mathfrak g^H\cdot z$ inside $V^H$ is contained in the orthogonal space to $\mathfrak g\cdot z$ inside $V$.
\end{lemma}

\begin{proof}
Take $u\in V^H$ orthogonal to all $\xi\cdot x$ for all $\xi\in\mathfrak g^H$. We have
\begin{equation}
0=\langle \xi v\mid x\rangle = \operatorname{Tr}\bigl(\xi\,|v\rangle\langle x|\bigr).
\end{equation}
Now take any $\eta\in\mathfrak g$ and let $\pi:\mathfrak g\to\mathfrak g^H$ be the orthogonal projection with respect to the negative-definite inner product $\operatorname{Tr}(AB)$ on skew-symmetric operators.
Since $\pi$ is $H$-equivariant and $|v\rangle\langle x|$ is fixed by $H$, we have
\[
\operatorname{Tr}\bigl(\eta\,|v\rangle\langle x|\bigr)=\operatorname{Tr}\bigl(\pi(\eta)\,|v\rangle\langle x|\bigr)=0.
\]
\end{proof}

\begin{corollary}
    Any vector in $V$ which is fixed by $H$ is in the $G^H$-orbit of a vector in $V$ in normal form which is also fixed by $H$.
\end{corollary}

Applying  this to $V=\frak{su}(2)_A\otimes \frak{su}(2)_B $ and $H$ the order $2$ group generated by the flip of the $2$ qubits we get

\begin{prop}
    A mixed $2$-qubit state which is invariant under the qubit flip is $SU(2)$-conjugate  to a mixed state of the form
    \begin{equation}
\sigma=\sigma_A\otimes \sigma_B+\alpha X\otimes X+\beta Y\otimes Y+ \gamma Z\otimes Z.
\end{equation}
\end{prop}

Here $SU(2)$ is the diagonal subgroup inside $SU(2)\times SU(2)$, which we view as the global gauge group inside the local gauge group.
\vskip .112 in
Similarly we have

\begin{prop}
    A $2$-qudit pure state which is invariant under flip is $SU(d)$-equivalent to a pure  state of the form 
  \begin{equation}
      c_1|11\rangle+\cdots+c_d|dd\rangle  
  \end{equation}  for real numbers $c_j$. 
\end{prop}

\section{The case of $3$-qubits}

The relevant vector space for $3$-qubits is $V=\C^2\otimes\C^2\otimes\C^2$. The  group $G$  is the local gauge group $G=SU(2)^3$.
We will consider $3$-qubit states which are invariant under the group $H=S_3$ of permutations of the $3$ qubits. The subgroup $G^H$ is the global gauge group $SU(2)$ inside $G=SU(2)^3$. The subspace $V^H$ is the $4$-dimensional complex subspace of $V$ spanned by $|000\rangle$, $|111\rangle$, $u$ and $v$ where
\[
 u=|100\rangle+|010\rangle+|001\rangle,\qquad v=|011\rangle+|101\rangle+|110\rangle.
\]

We will pick the vector $|\phi\rangle=|000\rangle+a|111\rangle$ for a real constant $a$.

\begin{lemma}
The tangent space to the $SU(2)$-orbit $\mathcal O$ of $|\phi\rangle$ is spanned by the three vectors
\[
 i\bigl(|000\rangle-|111\rangle\bigr),\quad u-v,\quad i(u+v).
\]
The normal space to the orbit is spanned by the five vectors
\[
 |000\rangle,\quad |111\rangle,\quad i\bigl(a|000\rangle+|111\rangle\bigr) , \quad i(u-v),\quad u+v.
\]
\end{lemma}

\begin{prop}
Any pure $3$-qubit state which is invariant under permutation of the $3$ qubits is equivalent under the global gauge group $G^H\cong SU(2)$ to a state of the form
\begin{equation}
 p|000\rangle+q|111\rangle + ir\bigl(a|000\rangle+|111\rangle\bigr) + is(u-v) + t(u+v)
\end{equation}
for real constants $p,q,r,s,t$.
\end{prop}

We note that this is different from the very economical normal form given in \cite{ac} for all pure $3$-qubit states.

\section{Gradient flow and a Riccati type equation}

To find a point $z$ in the $G$-orbit of $y$ which belongs to the normal space to the orbit $\mathcal O=G\cdot y$, one looks for a minimum of the smooth function
\[
F(g)=\lVert g\cdot y-x\rVert^2
\]
on $G$. One can then follow the gradient flow of $F$. We first need to compute the gradient of $F$, call it $\nabla F$.

We will use the bi-invariant Riemannian metric on $G$ induced by the one on $SO(V)$, so that
\[
\langle \xi\mid\eta\rangle =-\operatorname{Tr}(\xi\eta).
\]
We have computed that
\[
\bigl\langle (\nabla F)(g)\mid dR_g(\xi)\bigr\rangle=-2\operatorname{Tr}\bigl(\xi\,|gy\rangle\langle x|\bigr),
\]
where $R_g:G\to G$ denotes right translation by $g$. Since the inner product on tangent vectors to $G$ is invariant under $dR_g$ it follows that
\begin{equation}
(\nabla F)(g)=2\,dR_g\,\Pi\bigl(|gy\rangle\langle x|\bigr),
\end{equation}
where $\Pi:\operatorname{End}(V)\to \mathfrak g$ is the orthogonal projection when the matrix algebra $\operatorname{End}(V)$ is equipped with the inner product $\operatorname{Tr}(AB)$.

This requires some explanation since the inner product $\operatorname{Tr}(AB)$ on $\operatorname{End}(V)$ is neither positive nor negative definite. However its restriction to the space $\mathfrak{so}(V)$ of skew-symmetric operators is negative-definite, and so is its restriction to $\mathfrak g$. That suffices for the orthogonal projection $\Pi:\operatorname{End}(V)\to \mathfrak g$ to be well-defined.

Thus we have

\begin{prop}
The gradient flow $g(t)$ of the function $F$ on $G$ solves the differential equation
\begin{equation}
\frac{dg}{dt}=\Pi\bigl(|g(t)y\rangle\langle x|\bigr)\cdot g(t).
\end{equation}
\end{prop}
As the  time  derivative of $g(t)$ is a matrix valued quadratic polynomial  function of $g(t)$, we may view this ODE as a variant of a matrix-valued Riccati differential equation. Hopefully some methods in Riccati theory will apply to this case.

\printbibliography
\end{document}